\documentclass[11pt,a4paper]{article}
\usepackage{jheppub}
\usepackage[makeroom]{cancel}
\def\d{{\textrm  d}}

\newcommand{\nn}{\nonumber}

\newcommand{\Mbos}{{\cal M}}

\newcommand{\beq}{\begin{equation}}
\newcommand{\eeq}{\end{equation}}

\newcommand{\bea}{\begin{eqnarray}}
\newcommand{\eea}{\end{eqnarray}}

\title{Fermionic and bosonic mass deformations of ${\mathcal N}=4$ SYM and their bulk supergravity dual}

 \author[a]{Iosif Bena,} \author[a]{Mariana Gra\~na,}  \author[a]{Stanislav Kuperstein,} \author[a]{Praxitelis Ntokos} \author[b]{and Michela Petrini}

 \affiliation[a]{Institut de Physique Th\'eorique, Universit\'e Paris Saclay, CEA, CNRS, F-91191 Gif-sur-Yvette, France} 

 \affiliation[b]{LPTHE, Universit\'es Paris VI et VII, 4 place Jussieu 
 75252 Paris, France} 

\emailAdd{iosif.bena@cea.fr}\emailAdd{mariana.grana@cea.fr}\emailAdd{stanislav.kuperstein@gmail.com}\emailAdd{praxitelis.ntokos@cea.fr}\emailAdd{petrini@lpthe.jussieu.fr}

\abstract{We examine the AdS-CFT dual of arbitrary (non)supersymmetric fermionic mass deformations of $\mathcal{N}=4$ SYM, and investigate how the backreaction of the RR and NS-NS two-form potentials dual to the fermion masses contribute to Coulomb-branch potential of D3 branes, which we interpret as the bulk boson mass matrix. Using representation-theory and supergravity arguments we show that the fermion masses completely determine the trace of this matrix, and that on the other hand its traceless components have to be turned on as non-normalizable modes. Our result resolves the tension between the belief that the AdS bulk dual of the trace of the boson mass matrix (which is not a chiral operator) is a stringy excitation with dimension of order $(g_s N)^{1/4}$ and the existence of non-stringy supergravity flows describing theories where this trace is nonzero, by showing that the stringy mode does not parameterize the sum of the squares of the boson masses but rather its departure from the trace of the square of the fermion mass matrix. Hence, asymptotically-AdS flows can only describe holographically theories where the sums of the squares of the bosonic and fermionic masses are equal, which is consistent with the weakly-coupled result that only such theories can have a conformal UV fixed point.}

\preprint{IPhT-t15/027}

\begin{document}
\maketitle

\setcounter{footnote}{0}
\setcounter{figure}{0}
\setcounter{equation}{0}

\section{Introduction}

The $\mathcal{N}=4$ SYM theory deformed with three chiral multiplet masses, known as the $\mathcal{N}=1^\star$ theory, is one of the most studied examples of supersymmetric confining gauge theory, as it shares some of the most interesting features of QCD: confinement, baryons and flux tubes. Furthermore, since this theory has a conformal UV fixed point, it can be put on the lattice much easier than other four-dimensional gauge theories that one studies using the $AdS$/CFT correspondence, and hence can serve as an important benchmark for lattice gauge theory calculations.

The AdS/CFT dual of this theory has been spelled out by Polchinski and Strassler \cite{Polchinski:2000uf}, who deformed $AdS_5 \times S^5$ with non-normalizable modes in the RR and NSNS three-form fluxes, corresponding to masses for the fermions in the three chiral multiplets. They argued that in the resulting geometry the D3 branes that source $AdS_5 \times S^5$ polarize via the Myers effect \cite{Myers:1999ps} into spherical shells with five-brane dipole charge, that are the holographic duals of the confining, screening and oblique vacua of the $\mathcal{N}=1^\star$ theory \cite{Donagi:1995cf}.

The solution for the non-normalizable modes corresponding to the fermion masses and the existence of supersymmetry was enough to allow the authors of \cite{Polchinski:2000uf} to determine the full polarization potential of the D3 branes and to read off certain aspects of their physics. More precisely, in the limit when the number of five-branes is small, the polarization potential of the D3 branes has three terms. The first term, proportional to the fourth power of the polarization radius, is a universal term that gives the difference between the mass of unpolarized D3 branes and the mass of a five-branes with all these D3 branes inside. The second term, proportional to the third power of the radius, represents roughly the polarization force that the RR and NSNS three-form perturbations exert on the five-brane shell. The third term, proportional to the square of the radius,  is the potential felt by a probe D3-brane along what used to be the Coulomb-branch of the undeformed theory. This term comes from the backreaction of the three-forms dual to fermion masses on the metric, dilaton and five-form. In \cite{Polchinski:2000uf} the value of this term was guessed by using supersymmetry to complete the squares in the polarization potential. When the masses of the fermions in all the three chiral multiplets are equal, the value of this term was computed directly in supergravity by Freedman and Minahan \cite{Freedman:2000xb} and found to be exactly the one guessed in \cite{Polchinski:2000uf}.

Our main goal is to study the non-supersymmetric version of the Polchinski-Strassler story, and in particular to spell out a method to determine completely the D3-brane Coulomb branch potential (or the quadratic term in the polarization potential) for the $\mathcal{N}=4$ SYM theory deformed with a generic supersymmetry-breaking combination of fermion and boson masses. Many of the issues in the problem we are adressing have been touched upon in previous explorations, but when one tries to bring these pieces of the puzzle together one seems to run into contradictions. We will try to explain how these contradictions are resolved, and give a clear picture of what happens in the supergravity dual of the mass-deformed $\mathcal{N}=4$ theory. 

\medskip 

As explained in \cite{Polchinski:2000uf}, a fermion mass deformation of the $\mathcal{N}=4$ SYM field theory, $\lambda^i M_{ij} \lambda^j$, corresponds in the bulk to a combination of RR and NSNS three-form field strengths with legs orthogonal to the directions of the field theory, that transforms in the $\mathbf{10}$ of the $SU(4)$ R-symmetry group. The complex conjugate of the fermion mass, $ M^\dagger $, corresponds to the complex conjugate combination transforming in the $\overline{\mathbf{10}}$. Since the dimension of these fields is 3, the normalizable and non-normalizable modes dual to them behave asymptotically as $r^{-3}$ and $r^{-1}$.

The boson mass deformation in the field theory, $\phi^a {\cal M}_{ab} \phi^b$, can be decomposed into a term proportional to the trace of ${\cal M}$, which is a singlet under the $SU(4)\simeq SO(6)$ R-symmetry, and a  symmetric traceless mass operator, which has dimension 2 and transforms in the $\mathbf{20}^\prime$ of $SO(6)$. The traceless mass operator in the  $\mathbf{20}^\prime$ corresponds in the  $AdS_5 \times S^5$ bulk dual to a deformation of the metric, dilaton and the RR four-form potential that is an $L=2$ mode on the five-sphere, and whose normalizable and non-normalizable asymptotic behaviors are $r^{-2}$ and $r^{-2} \log r$ \cite{Kim:1985ez}. On the other hand, the dimension of the trace operator is not protected, and hence, according to the standard lore, turning on this operator in the boundary theory does not correspond to deforming $AdS_5 \times S^5$ with a supergravity field\footnote{This is consistent with the fact that there are no perturbations around $AdS_5 \times S^5$ that are $SO(6)$ singlets and behave asymptotically as $r^{-2}$ and $r^{-2} \log r$.}, but rather with a stringy operator \cite{Witten:1998qj}. The anomalous dimension of this operator at strong coupling has consequently been argued to be of order $(g_s N)^{1/4}$.

On the other hand, there exist quite a few supergravity flows dual to field theories in which the sum of the squares of the masses of the bosons are not zero \cite{Freedman:1999gp,Girardello:1999bd,Pilch:2000ue,Evans:2000ap,Pilch:2000fu,Khavaev:2000gb,Bigazzi:2003ui,Apreda:2003sy,Gowdigere:2005wq}, and none of these solutions has any stringy mode turned on, which seems to contradict the standard lore above. In this paper we would like to argue that the solution to this puzzle comes from the fact that 
the backreaction of the bulk fields dual to the fermions determines completely the singlet piece in the quadratic term of the Coulomb branch potential of a probe D3-brane. Therefore, the trace of the boson mass matrix  that one reads off from the bulk will always be equal to the trace of the square of the fermion mass matrix. 

This, in turn, indicates that in the presence of fermion masses, the stringy operator is not dual to the sum of the squares of the boson masses, but to the difference between it and the sum of the squares of the fermion masses.  Mass deformations of the $\mathcal{N}=4$ theory where the supertrace of the square of the masses is zero can therefore be described holographically by asymptotically-$AdS$ supergravity solutions \cite{Freedman:1999gp,Girardello:1999bd,Pilch:2000ue,Evans:2000ap,Pilch:2000fu,Khavaev:2000gb,Bigazzi:2003ui,Apreda:2003sy,Gowdigere:2005wq}. However, to describe theories where this supertrace is nonzero, one has to turn on ``stringy'' non-normalizable modes that correspond to dimension-$(g_s N)^{1/4}$ operators, which will destroy the $AdS$ asymptotics.

To see this we begin by considering the backreaction of the three-form field strengths corresponding to fermion mass deformations on the metric, the dilaton and the four-form potential, which has been done explicitly for several particular choices of masses \cite{Freedman:2000xb, Taylor:2001pp}. This backreaction can give several terms that modify the action of a probe D3 brane, giving rise to a Coulomb-branch potential that is quadratic in the fermion masses and that transforms either in the $\mathbf{1}$ or in the $\mathbf{20}^\prime$ of $SO(6)$. Furthermore, one can independently turn on non-normalizable modes in the $\mathbf{20}^\prime$ of $SO(6)$ that correspond to deforming the Lagrangian with traceless boson bilinears, and that can also give rise to a Coulomb-branch potential. Since all these terms behave asymptotically as $r^{-2}$ and transform in the same $SO(6)$ representation, disentangling the contributions of the non-normalizable modes from the terms coming from the backreaction of the three-forms can be quite nontrivial. For example, in equation (62) in \cite{Polchinski:2000uf}, the Coulomb-branch potential appears to contain both contributions in the $\mathbf{1}$ and in the $\mathbf{20}^\prime$ of $SO(6)$ coming from the backreaction of the fermion mass tensor $T_{ijk}$, and to have no non-normalizable contribution.

We will show that the backreaction of the modes dual to the fermion masses can only source terms in the D3 brane Coulomb-branch potential that are singlets under $SO(6)$, and hence the Coulomb-branch potential terms that transform in the $\mathbf{20}^\prime$ of $SO(6)$ can only come from non-normalizable $L=2$ (traceless) modes that one has to turn on separately from the fermion masses. Since the singlet term in the Coulomb-branch potential is the supergravity incarnation of the trace of the boson mass matrix, our result implies that in the bulk this boson mass trace is completely determined by the fermion masses: the sum of the squares of the boson masses will always be equal to the sum of the squares of the fermion masses.

Our calculation establishes that asymptotically-$AdS_5$ solutions can only be dual to theories in which the sum of the squares of the boson masses is the same as the sum of the squares of the fermion masses. Theories where these quantities are not equal cannot by described holographically by such solutions.

From a field theory perspective this interpretation is very natural: the solutions that are asymptotically $AdS_5$ can only be dual to field theories that have a UV conformal fixed point, and therefore their masses and coupling constants should not run logarithmically in the UV (their beta-functions should be zero). At one loop this cannot happen unless the sum of the squares of the boson masses is equal to the sum of the squares of the fermion masses \cite{usshort}, which reduces the degree of divergence in the corresponding Feynman diagram and makes the beta-functions vanish.\footnote{Note that this discussion only applies to asymptotically-$AdS_5$ backgrounds. The Klebanov-Strassler solution \cite{Klebanov:2000hb}, which is not asymptotically-$AdS_5$, is dual to a field theory where the coupling constants run logarithmically. } Thus in perturbative field theory one inputs boson and fermion masses, and one cannot obtain a 
 UV conformal fixed point unless the sums of their squares are equal; in contrast,  in holography one inputs an asymptotically-AdS solution (dual to a conformal fixed point) and the non-normalizable modes corresponding to fermion masses, and obtains automatically the sum of the squares of the boson masses.

This understanding of how the sum of the squares of the boson masses appears in AdS-CFT also clarifies some hitherto unexplained miraculous cancellations. In the Pilch-Warner dual of the ${\cal N}=2^\star$ theory \cite{Pilch:2000ue}, which from the ${\cal N}=1$ perspective has a massless chiral multiplet and two chiral multiplets with equal masses, the only non-normalizable modes that were turned on in the UV were those corresponding to the fermion masses $M={\rm diag}(m,m,0,0)$ and to a traceless ($L=2)$ boson bilinear of the form  ${m^2 \over 3} (|\phi_1|^2 + |\phi_2|^2 - 2 |\phi_3|^2)$. Since the latter contains some tachyonic pieces one could have expected the potential for the field $\phi_3$ to be negative, but in the full solution this potential came out to be exactly zero. Using the new understanding developed in this paper it is clear that this ``miraculous cancellation'' happens because the backreaction of the fields dual to fermion masses gives a non-trivial contribution to the trace of the boson mass, of the form ${ 2 m^2 \over 3} (|\phi_1|^2 + |\phi_2|^2 + |\phi_3|^2)$, and as a result the potential for $\phi_3$ exactly cancels. 

One of the motivations for our work is the realization that the near-horizon regions of anti-branes in backgrounds with charges dissolved in fluxes have tachyonic instabilities \cite{Bena:2014bxa, Bena:2014jaa}. From the point of view of the $AdS$ throat sourced by the anti-branes, this tachyon comes from a particular $L=2$ bosonic mass term that is determined by the gluing of this throat to the surrounding region. Understanding the interplay between this mass mode and the fluxes of the near-brane region is crucial if one is to determine whether the tachyonic throat has any chance of supporting metastable polarized brane configurations of the type considered in the KPV probe analysis \cite{Kachru:2002gs}. Preliminary results of this investigation have already appeared in \cite{Bena:2015kia}.

The paper is organized as follows. In Sections \ref{sec:Group-Theory} and \ref{sec:The-Explicit-Map} we use group theory to find the bosonic potential, both the singlet and the $\mathbf{20}^\prime$ pieces, arising from the square of the fermionic masses living in the $\mathbf{10}$ of $SU(4)$. Although the group theory is  well-known and most of Section \ref{sec:Group-Theory} is a review, our final formulas in Section \ref{sec:The-Explicit-Map} are new, as only their supersymmetric versions have so far appeared in the literature. In Section \ref{sec:The-Mass-Deformation} we explain how the bosonic masses appear in supergravity. This section contains the main observations of the paper. In Section \ref{sec:Future} we recapitulate the main conclusions of our analysis and their relation to perturbative gauge theories.  The appendix includes a summary of useful formulas for intertwining between $SO(6)$ and $SU(4)$ representations.

\section{The  Group Theory of the Mass Deformations}
\label{sec:Group-Theory}

The goal of this section is to identify the $SO(6)$ representation of the fermionic and bosonic mass deformations. We begin by reviewing in detail the group theory behind the mass deformations because this will play an important role in our discussion. 

\subsection{Fermionic masses}

The most general non-supersymmetric fermionic mass deformation of $\mathcal{N}=4$ SYM is given by the operator:\footnote{We use $i,j,k, \ldots=1,2,3,4$ indices for the fermions (i.e. for the fundamental of $SU(4)$) and $A,B,C, \ldots=1, \ldots, 6$ for the bosons (fundamental $SO(6)$ representation).}
\begin{equation}
\lambda^i M_{ij} \lambda^j \, ,
\end{equation}
where $\lambda^i, i=1,...,4$ are the 4 Weyl fermions of the ${\cal N}=4$ theory, that in ${\cal N}=1$ language are the three fermions in chiral multiplets plus the gaugino. The mass matrix $M$ is in the ${\bf 10}$ of SU(4), which is the symmetric part of  $4 \times 4$:
\begin{equation} \label{4x4}
\textbf{4} \times \textbf{4} = \textbf{6}_{\textrm {a}}+ \textbf{10}_{\textrm {s}} \, .
\end{equation}
As noted in \cite{Polchinski:2000uf}, this matrix in the ${\bf 10}$ of $SU(4)\cong SO(6)$ can equivalently be encoded in an imaginary anti-self dual 3-form\footnote{In our conventions the anti-self duality means $(\star_6 T)_{ABC}=\frac{1}{ 3!} \epsilon_{A B C}{}^{ D E F} T_{D E F} = - i T^{A B C}$.} $T_{ABC}$. The map between them will be given in the next section.

In the language of ${\cal N}=1$, one distinguishes a $U(1)_R \subset SU(4)_R$ that singles out the gaugino within the 4 fermions, or in other words the $SU(4)$ $R$-symmetry group is broken as:
\begin{equation} \label{SU4SU3}
SU(4)_R \to SU(3) \times U(1)_R \, 
 \end{equation}  
corresponding to the splitting of the fundamental index ${\bf 4}={\bf 3}+{\bf 1}$ ($i=\{I,4\}$). In this breaking, the fermionic mass matrix in the ${\bf 10}$ decomposes as
\begin{equation}
{\bf 10}={\bf 6}+ {\bf 3}+ {\bf 1} \ .
\end{equation}
This corresponds to the breaking of $M$ into the following pieces
\begin{equation} \label{MSU3}
 M_{ij} = \begin{pmatrix} m_{IJ} &  \hat m_I \\ \hat m_{I}^T &  \tilde m \end{pmatrix} \ 
 \end{equation}
where $m_{IJ}$, $\hat m_I$ and $\tilde m$ are respectively in the ${\bf 6}$, ${\bf  3}$ and ${\bf 1}$.

\subsection{Bosonic Masses}

A generic $6 \times 6$ bosonic mass matrix ${\cal M}^2_{AB}$ has 21 components, coming from the symmetric piece in
\begin{equation} \label{6x6}
(\mathbf{6} \times {\mathbf{6}})_s = \mathbf{1} + \mathbf{20'} \ . 
\end{equation}

If bosonic masses come from the backreaction of the fermion masses on the supergravity fields, ${\Mbos}^2$ should be of order $M^2$. The most naive guess is that they are related to the hermitian matrix $M M^\dagger$, which involves the following SU(4) representations:
\begin{equation}
\label{eq:10-times-10-bar}
\textbf{10} \times \overline{\textbf{10}} = \textbf{1} + \textbf{15} + \textbf{84} \, .
\end{equation}

From these very simple group-theory arguments one can immediately conclude that either our naive guess was too simple, or that the backreaction of the fermionic masses only generates the singlet (the trace) in the bosonic masses. However, since this goes against most people's intuition, particularly when there is some supersymmetry preserved, let us then push a bit further the possibility that our naive guess was wrong, or in other words that the bosonic masses are determined by fermionic ones, and see where it takes us. 

The ${\bf 20}'$ representation in \eqref{6x6}, which is not in the product \eqref{eq:10-times-10-bar}, appears instead in 
\begin{equation}
\label{eq:10-times-10}
\textbf{10} \times \textbf{10} = \textbf{20}^\prime_{\textrm {s}} + \textbf{35}_{\textrm {s}} + \textbf{45}_{\textrm {a}} \, . 
\end{equation}
In terms of $SU(4)$, the $\textbf{20}^\prime_{\textrm {s}}$ is one of the three $20$-dimensional representations whose Young tableau and Dynkin label are:
\begin{equation}
\label{eq:20}
\textbf{20}^\prime = \begin{array}{c} \centering \textrm{\Huge{$\boxplus$}} 
\end{array}
= \big( 0\,2\,0 \big)
\, .
\end{equation}
There is an important caveat here: this representation is complex, and we therefore have to project out half of the components in order to get a real representation for the bosonic masses. As we will see in the next section this projection is directly related to the map between $SU(4)$ and $SO(6)$.
A straightforward check that this representation is the one describing bosonic masses is to see what happens when ${\cal N}=1$ supersymmetry is preserved ($\hat m_I=\tilde m=0$ in \eqref{MSU3}). The bosonic mass matrix should then be proportional to $m m^{\dagger}$ in 
\begin{equation}
\mathbf{3} \times \overline{\mathbf{3}} = \mathbf{1} + \mathbf{8}  \, .
\label{susymass}
\end{equation}
The ${\bf 1}$ representation is the one we discussed above, while the ${\bf 8}$ representation indeed appears in ${\bf 20}'$, with the right $U(1)_R$ charge, since   
for the breaking \eqref{SU4SU3}, we have \cite{Slansky:1981yr}:
\begin{equation}
\textbf{20}^\prime = \overline{\textbf{6}} (-4/3) + \textbf{6}(4/3) + \textbf{8}(0) \, .
\end{equation}

From these group-theory arguments we conclude that if boson masses are generated by fermion masses at second order, then
\begin{equation}
\textrm{Tr} \left( MM^\dagger \right) \to \textrm{Tr} \left( \Mbos^2 \right)
\end{equation}
while the other 20 components of ${\Mbos}^2$ come from the product $M M$. Anticipating, we will see this map explicitly in the next section, from which we will   
conclude that only the former is true.

\section{The explicit map between bosonic and fermionic mass matrices}
\label{sec:The-Explicit-Map}

In this section we will construct explicitly the maps (\ref{eq:10-times-10-bar}) and (\ref{eq:10-times-10}), and the relationship between $SU(4)$ and $SO(6)$ representations. This will give the  form of the possible terms in the supergravity fields that depend quadratically on fermion masses, which come from the backreaction of the fields dual to these masses. As shown in the previous section, the backreaction splits into two parts, corresponding to the $\mathbf{20}^\prime$ and $\mathbf{1}$ representations.

To build a map between $SU(4)$ and $SO(6)$ one identifies the $\mathbf{6}_\textbf{a}$ representation of $SU(4)$ we have encountered above in \eqref{4x4} with the fundamental representation of $SO(6)$. The former is given by a $4 \times 4$ antisymmetric matrix, $\varphi^\textrm{T}=-\varphi$, that transforms as $\varphi \to U \varphi U^\textrm{T}$ under $U \in SU(4)$. The complex $\mathbf{6}$ can be further decomposed into two real representations, $\mathbf{6}=\mathbf{6}_+ + \mathbf{6}_-$, by imposing the duality condition:\footnote{The projection commutes with $SU(4)$ since $\epsilon^{ijkl}$ is an invariant tensor.}
\begin{equation}
\label{eq:6-projection}
\star\varphi=\pm \varphi^{\dagger} \ , 
\end{equation}
where $(\star\varphi)^{ij}=\frac{1}{2} \epsilon^{ijkl} \varphi_{kl}$. 
In what follows we will use the following parametrization of $\mathbf{6}_+$:
\begin{equation}
\label{eq:6-parametrization}
\varphi = \left(
\begin{array}{cccc}
0 & \bar{\varPhi}_3 & -\bar{\varPhi}_2 & -\varPhi_1 \\
- \bar{\varPhi}_3 & 0 &  \bar{\varPhi}_1 & -\varPhi_2 \\
\bar{\varPhi}_2 & -\bar{\varPhi}_1 & 0 & -\varPhi_3 \\
\varPhi_1 & \varPhi_2 & \varPhi_3 & 0 
\end{array}
\right) \, ,
\end{equation}
where the $\varPhi_{1,2,3}$ are complex combinations of the six real scalars $\phi^{A=1,\ldots,6}$ in the fundamental representation of $SO(6)$. We choose conventions such that $\varPhi_I = \phi^I + i \, \phi^{I+3}$ for $I=1,2,3$. This parametrization is convenient as it makes explicit the $\mathbf{6} \to \mathbf{3} +  \overline{\mathbf{3}}$ decomposition and the relation with the three chiral multiplets of $\mathcal{N}=4$. From (\ref{eq:6-parametrization}) we find:
\begin{equation}
\label{eq:6-map}
\varphi_{ij} = \sum_{A=1}^6 {G^A}_{ij} \phi^A 
\qquad
\textrm{or}
\qquad
\phi^A = \frac14 {{G^A}}^{ij} \varphi_{ji}
\, ,
\end{equation}
where the six matrices $G^A$ are antisymmetric self-dual matrices (sometimes referred as 't Hooft symbols, or generalized Weyl matrices) which intertwine between $SO(6)$ and $SU(4)$, and whose form and explicit properties we give in Appendix \ref{sec:t-Hooft-matrices}, and  ${G^A}^{ij} \equiv \overline{G}^A_{ji}$. An $SU(4)$ rotation given by a matrix $U$ is related to an $SO(6)$ rotation by a matrix $O$ via:\footnote{The $SO(6)$ indices are raised with $\delta^{AB}$.}
\begin{equation}
  U^{\, k}_i {G^A}_{kl} U^{\,l}_{ j}=O^A_{\, B} {G^B}_{ij} \qquad \textrm{or} \qquad 
  O^{A B}\equiv \dfrac{1}{4}  G^A_{kl} U_{ j}^{\,l} {G^B}^{ji} U_{i}^{\, k} \, .
\end{equation}
Note that the action of $SO(6)$ is the same when $U \to -U$, and so, as expected, $SO(6)=SU(4)/\mathbb{Z}_2$.

With the help of t'Hooft matrices, we can work out the explicit map between the fermion mass matrix $M_{ij}$ and an anti-self dual 3-form $T_{ABC}$. We get
\begin{equation} 
\label{TM}
T_{A B C} = -\frac{1}{2 \sqrt{2}} \textrm{Tr} \left( M G^A {G^B}^{\dagger} G^C \right) \ , \quad M_{ij}= \frac{1}{12 \sqrt{2}} T_{ABC} ({G^A}^{\dagger}  {G^B} {G^C}^{\dagger} )_{ij} \ ,
\end{equation}
where the trace in the first expression is over the $SU(4)$ indices and the numeric factors are chosen to reproduce (35) of \cite{Polchinski:2000uf} for a diagonal $M$. One can use the properties of the 't Hooft matrices in (\ref{eq:G-properties-1}) and  (\ref{eq:G-properties-2}) to verify that $T_{A B C}$ is indeed an anti-self-dual three-form. 

In terms of the 3-form $T$, the different representations correspond to the following components:\footnote{The primitive ${\bf 6}$ and non-primitive ${\bf \bar 3}$ pieces of a 3-form $G$ are obtained as follows
\begin{eqnarray}
 G^{\bf 6}_{I\bar J \bar K}&=&G_{I\bar J \bar K}- J_{I[\bar J} \, G_{\bar K] L \bar M} J^{L \bar M} \nn \\  
 G^{\bf \bar 3}_{I \bar J \bar K}&=& J_{I [\bar J} \,  G_{\bar K] L \bar M } J^{L \bar M} \ . \nn 
\end{eqnarray} }
\begin{eqnarray} \label{TandM}
{\bf 6} : (1,2) \, \text{primitive} \quad T_{I\bar J \bar K}=T^{\bf 6}_{I\bar J \bar K} \ &,&  \quad  \tfrac12 T_{I\bar J \bar K} \epsilon^{\bar J \bar K}{}_L =  m_{IL}  \nn \\  
{\bf 3} :(2,1) \, \text{non-primitive} \quad T_{I J \bar K}=T^{\bf 3}_{I J \bar K}\ &,&  \quad  \tfrac{i}{2} T_{I J \bar K} J^{J \bar K}= -\hat m_I  \\  
{\bf 1} : (3,0) \quad T_{I J  K}= T^{\bf 1}_{I J  K} \quad &,& \quad  \tfrac16 T_{I J  K} \epsilon^{IJK} =  \tilde m \nn 
\end{eqnarray}
where $J_{I\bar J}$ is the symplectic structure associated to the $SU(3)$ group. In our conventions it is just $
J_{1 \bar 1}=J_{2 \bar 2}=J_{3 \bar 3}=i$.

Let us now discuss the bosonic masses, in the ${\bf 20}'+ {\bf 1}$ representations of $SO(6)$. In terms of $SU(4)$, the $\mathbf{20}^\prime$ representation is labelled by four indices and from its Young tableau  (\ref{eq:20}) we learn that:
\begin{equation} \label{20p}
B_{ij,kl} = B_{kl,ij} = - B_{ji,kl} = -B_{ij,lk} \, . 
\end{equation} 
Furthermore, the zero-trace condition
\begin{equation}
\label{eq:20-tracelessness}
\epsilon^{ijkl} B_{ij,kl} = 0
\end{equation}
eliminates the singlet leaving only $\mathbf{20}^\prime$ from $\mathbf{20}^\prime \oplus \mathbf{1}$. Following our discussion we can decompose this complex $SU(4)$ representation into two real $SO(6)$ representations, $\mathbf{20}^\prime_\mathbb{C} = \mathbf{20}^\prime_+ + \mathbf{20}^\prime_-$.
This is achieved by requiring:
\begin{equation}
\label{eq:20-projection}
\overline{B_{ij,kl}} = \pm \frac{1}{4} \epsilon^{ijmn} B_{mn,pq}\epsilon^{pqkl} \, ,
\end{equation}
and we will use in this paper the choice $\mathbf{20}^\prime_+$.
The explicit map between the $\mathbf{20}^\prime$ representations of $SU(4)$ and $SO(6)$ then works very similarly to (\ref{eq:6-map}):
\begin{equation}
\label{eq:V-B}
V_{\mathbf{20}^\prime}^{A B} = \frac{1}{4} {G^A}^{ij} B_{ij,kl} {G^B}^{kl} \, .
\end{equation}
It is straightforward to verify that $V_{\mathbf{20}^\prime}^{A B}$ is symmetric and real when $B_{ij,kl}$ satisfies (\ref{20p}) and (\ref{eq:20-projection}) with the upper sign. Moreover, by using the fact that the 't Hooft matrices satisfy \eqref{eq:G-properties-1}, 
one can see that the tracelessness of $V_{\mathbf{20}^\prime}^{A B}$ is guaranteed by (\ref{eq:20-tracelessness}).

Now, given a fermionic mass matrix $M$, one  can build the following matrix in the ${\bf 20}'_+$ : 
\begin{equation}
\label{eq:B-from-mm}
B_{ij,kl} =  \frac{1}{2} \left( M_{ik} M_{jl}-  M_{il} M_{jk} \right) + \frac{1}{4} \epsilon_{ijpq} \epsilon_{rskl} \overline{M}^{pr} \overline{M}^{qs} \, .
\end{equation}
Here the first term is dictated by the Young tableau (\ref{eq:20}) and the second guarantees (\ref{eq:20-projection}) with the $\textbf{20}^\prime_+$ choice. 
Furthermore, it is by construction traceless. One can add a trace to this, which, as discussed, should be built from $MM^\dagger$. We define
\begin{equation}
\widetilde{B}_{ij,kl} = -\frac{1}{2} \epsilon_{ijkl} \textrm{Tr} \left( M M^\dagger \right) \, ,
\end{equation}
 which in turn, using the properties listed in the Appendix, implies that:
\begin{equation}
\label{eq:tilde-V-mm-bar}
V_{\mathbf{1}}^{A B} \equiv   \frac{1}{4} {G^A}^{ij} \tilde B_{ij,kl} {G^B}^{kl}=  \textrm{Tr} \left( M M^\dagger \right) \delta^{A B}  \, .
\end{equation}

To summarize, the most general bosonic mass matrix produced by the backreaction of the fermionic masses is $V^{AB}_{\rm quad.}$, given by some linear combination of the $\textbf{20}^\prime$ and $\textbf{1}$ contributions, $V_{\mathbf{20}^\prime}^{A B}$ and $V_{\mathbf{1}}^{A B}$. The latter is related to the fermion masses as in (\ref{eq:tilde-V-mm-bar}), while the former is determined by (\ref{eq:V-B}) with (\ref{eq:B-from-mm}). Out of this we can build a scalar $\phi^A V^{AB}_{\rm quad.} \phi^B$, or identifying the scalars $\phi^A$ with some local coordinates on the six-dimensional space $x_A$ we get the ``potentials"
\begin{equation} \label{V1V20}
 V_{\mathbf{1}} \equiv x_A V_{\mathbf{1}}^{AB} x_B  \ , \qquad  V_{\mathbf{20}^\prime} \equiv x_A V_{\mathbf{20}^\prime}^{A B} x_B \ .
\end{equation}
Let us now examine the form of these potentials for the simple example of a diagonal fermionic mass matrix:
\begin{equation}
\label{fermionicmassmatrix}
M = \textrm{diag} \left( m_1, m_2, m_3, m_4 \right) \, ,
\end{equation}
which yields
\begin{eqnarray} 
\label{pot20}
V_{\mathbf{1}}~~ &=& (|m_1|^2+|m_2|^2+|m_3|^2+|m_4|^2) \, \left(x_1^2+ \ldots +x_6^2 \right)  \,   \\
V_{\mathbf{20}^\prime}&=& {\rm Re}(m_2 m_3+m_1 m_4) (x_1^2 - x_4^2) +{\textrm Re} (m_1 m_3+m_2 m_4) (x_2^2-x_5^2) \nonumber \\
&+& {\rm Re} (m_1 m_2 + m_3 m_4) (x_3^2-x_6^2)  -2\, {\rm Im} (m_2m_3-m_1m_4)\, x_1 x_4
\nonumber \\
&-& 2\, {\rm Im} (m_1m_3-m_2m_4)\, x_2 x_5  -2 \,{\rm Im} (m_1m_2-m_3m_4) \, x_3 x_6  \ . \nonumber
\end{eqnarray}
It is not hard to see that when the fourth fermionic mass is zero, and hence ${\cal N}=1$ supersymmetry is preserved, there is no combination of these two terms that can yield the $\mathcal{N}=1^\star$ supersymmetric bosonic mass potential

\begin{equation}
\label{eq:V-SUSY}
V_{\mathcal{N}=1^\star}  = \left| m_1 \right|^2 \left( x_1^2 + x_4^2 \right) + \left| m_2 \right|^2  \left( x_2^2 + x_5^2 \right) + \left| m_3 \right|^2  \left( x_3^2 + x_6^2 \right) \, .
\end{equation}
 Hence, the bosonic mass matrix cannot be fully determined by the fermion mass matrix.

\section{The mass deformation from supergravity}
\label{sec:The-Mass-Deformation}

In this section, we will discuss how to get the bulk boson masses from the dual supergravity solution given by the full backreaction of the dual of the fermion masses on $AdS_5 \times S^5$. The fully backreacted ten-dimensional (Einstein frame) metric is generically of the form
\begin{equation}
ds^2=e^{2A} \eta_{\mu \nu} dy^\mu dy^\nu + ds_6^2 \ ,
\end{equation}
with the RR four-form potential along space-time
\begin{equation}
C_4=\alpha \,  \d y^0 \wedge \ldots \wedge \d y^3 \ , 
\end{equation}
a dilaton $\phi$ and some internal 3-form fluxes that are usually combined into the complex form
\begin{equation}
G_3=F_3-\tau H_3 \ ,
\end{equation}
where $\tau=C - i e^{-\phi}$ is the combination of RR axion and dilaton.

As explained in the Introduction, and as can be seen from the explicit flow solutions corresponding to mass deformations of 
$\mathcal{N}=4$ theory that have been constructed explicitly \cite{Freedman:1999gp,Pilch:2000ue,Girardello:1999bd} the boson masses  can be read off from the quadratic terms in the D3 Coulomb-branch potential, given by:
\begin{equation} \label{VD3}
V_\textrm{D3}=\int d^4 y \sqrt{g_\parallel} - \int C_4=\int d^4 y \,  (e^{4A}-\alpha)  \ ,
\end{equation}
where the warp factor and four-form potential are those of the fully backreacted solution. This computation is quite complicated for generic fermion masses, and was only obtained for some special choices, corresponding to the equal-mass ${\cal N}=1^{\star}$ theory ( $M={\rm diag}(m,m,m,0)$) \cite{Freedman:2000xb} and the supersymmetry-breaking-$SO(4)$-invariant ${\cal N}=0^{\star}$ theory ($M={\rm diag}(m,m,m,m)$)  \cite{Taylor:2001pp}.  We will see how much of the quadratic term of $V$ we can infer from these examples and from our group-theoretic arguments in the previous sections.

On the gravity side the fermionic mass deformation corresponds to the non-normalizable modes of the complex 3-form flux $G_3$  \cite{Girardello:1999bd},\cite{Polchinski:2000uf}. As we argued in the previous sections, the  ${\bf 10}$ representation of the SU(4) fermion mass matrix $M_{ij}$ is equivalent to the ${\bf 10}$ of SO(6) corresponding to imaginary anti-self-dual 3-forms.    
At first order in the mass perturbation the supergravity equations of motion are satisfied if the imaginary anti-self-dual 3-form $e^{4A}( \star_6 G_3-i G_3)$ is closed and co-closed. One option is to set this to zero, i.e. to have $G_3$ be purely in the $\overline {\bf 10}$ (imaginary self-dual), but this solution does not correspond to the dual of the ${\cal N}=1^\star$ gauge theory.\footnote{On this solution, the D3-branes feel no force, which implies that the potential is zero.} The three form flux has therefore both ${\bf 10}$ and ${\bf \overline{10}}$ components, and has the $r^{-1}$ behavior of a non-normalizable mode dual to the $\Delta=3$ operator corresponding to the fermion masses. It is given by:
\begin{equation} \label{3form}
G_3=\frac{c}{r^4} \left(T_3-\frac{4}{3} V_3 \right) \, ,
\end{equation}
where $c$ is a constant, $T_3$ is the imaginary anti-self-dual 3-form corresponding to the fermion masses, Eq. (\ref{TM}), and $V_3$ is constructed from $T_{3}$ and combinations of the vector $x^A$, and it has both $\mathbf{10}$ and $\overline{\mathbf{10}}$ components:
\begin{equation} \label{V}
V_{A B C} = \frac{3}{r^2} x^D x_{[ A} T_{B C ] D} \ .
\end{equation}

At second order (quadratic in the fermionic masses) one has to solve for the dilaton, the metric and the 4-form potential, whose equations of motion depend quadratically on $G_3$, and this was only done for the special mass deformations discussed above  \cite{Freedman:2000xb,Taylor:2001pp};  for supersymmetric unequal masses only the solution for the dilaton-axion is known \cite{Grana:2000jj}.  Here we will not need the details of these solutions, but we note a few key points from which we will draw our conclusions. 

 The EOMs for the dilaton, warp factor and four-form potential have schematically the following structure:
\begin{equation}
\label{eq:B-F2}
\vec \nabla \cdot \vec \nabla \left( \textrm{Bosonic fields} \right) = \left( \textrm{3-form Fluxes} \right)^2 \, ,
\end{equation}
Since the fluxes are known, a general solution for the bosonic fields has inhomogeneous and homogeneous parts. 

For fluctuations around $AdS_5\times S^5$, the homogenous part is a combination of harmonics of the sphere with different fall-offs in $r$. The quadratic term in \eqref{VD3} comes from modes with a $r^{-2}$ fall off (the background warp factor $e^{4A_0}\sim r^4$), or in other words from modes which are dual to an operator of  dimension $\Delta=2$. Only the ${\bf 20'}$ representation in the combination of metric and four-form potential that is relevant to compute (\ref{VD3}) has this behavior \cite{Kim:1985ez}. It corresponds to the second harmonic on the five-sphere, and was referred in \cite{Polchinski:2000uf} as the $L=2$ mode. 

The inhomogeneous piece is sourced by quadratic combinations of the three-form fluxes, which transform in (\ref{eq:10-times-10-bar}) and (\ref{eq:10-times-10}). Out of these, only the 
${\bf 1}$ and ${\bf 20'}$ contribute to the masses of the bosons.
%
%
The corresponding pieces in the fields that give rise to these masses can then be schematically represented as:
\begin{eqnarray}
\label{eq:general-solution}
 \phi & \sim & f^{\phi }_\textrm{inhom.} (r) V_{\mathbf{20}^\prime} +  g^{\phi }_\textrm{inhom.} (r) V_{\mathbf{1}} + h^{\phi}_\textrm{hom.} (r)
\, U_{\mathbf{20}^\prime}  \, \nonumber \\
g_{\parallel} & \sim & f^{g}_\textrm{inhom.} (r) V_{\mathbf{20}^\prime} +  g^{g}_\textrm{inhom.} (r) V_{\mathbf{1}} + h^{g}_\textrm{hom.} (r)
\, U_{\mathbf{20}^\prime}  \, \\
 \alpha & \sim & f^{RR}_\textrm{inhom.} (r) V_{\mathbf{20}^\prime} +  g^{RR}_\textrm{inhom.} (r) V_{\mathbf{1}} + h^{RR}_\textrm{hom.} (r)
\, U_{\mathbf{20}^\prime} \, , \nonumber
\end{eqnarray}
where the first two terms in each line correspond to inhomogeneous solutions, whose dependence on the fermionic mass we computed in the previous section (equation (\ref{pot20}) for a diagonal mass matrix), and the last term is the contribution from the homogeneous solution whose angular dependence,
\begin{equation}
\label{U20def}
U_{\mathbf{20}^\prime} \equiv   x^A \mu_{AB}^{\mathbf{20}^\prime} x^B  \, , 
\end{equation}
is determined by 20 free parameters $\mu_{AB}^{\mathbf{20}^\prime}$, 
that have the dimension of mass squared.\footnote{Only a subset of these are possible in a symmetric configuration. For example, when an $SO(3)$ symmetry is preserved ($M={\rm diag}(m,m,m,\tilde m)$), there are only two invariant parameters \cite{Zamora:2000ha}.} It is important to note that, unlike the components of $V_{\mathbf{20}^\prime}$, the components  of $U_{\mathbf{20}^\prime}$ are \emph{not} related in any direct way to the fermionic masses $M_{ij}$, but are determined in a given configuration by IR and UV boundary conditions.  

With the solution for the metric and the 4-form potential at hand, one can compute the boson masses  directly in supergravity, through (\ref{VD3}). If one works in Einstein frame, this requires only the combination of warp factor and four-form potential $\Phi_-=e^{4A}-\alpha$, whose equation of motion has a right-hand side of the form (see (2.30) of \cite{Giddings:2001yu}):
\begin{equation}
\label{eq:BoxPhi-}
\Box \left( \Phi_- \right)\propto \left\vert \star_6 G_3-i G_3 \right\vert^2 + \ldots  \propto \left\vert T_3 \right\vert^2 + \ldots  \, ,
\end{equation} 
where the $\ldots$ stand for the terms that are higher order in the mass deformation, and  in the last step we have used (\ref{3form}) together with the duality properties $ \star_6 T_3=-i T_3$ and $ \star_6 V_3=-i \left( T_3-V_3 \right)$. The crucial observation is that $V_3$ drops out of the equation. The remaining piece,  $\left\vert T_3 \right\vert^2$, has no $x$-dependence and as a result is proportional to the singlet of the $\mathbf{10} \times \overline{\mathbf{10}}$ product.
We see that out of the $\mathbf{20}^\prime$ and the $\mathbf{1}$ parts in the inhomogeneous solution (\ref{eq:general-solution}), only the latter contributes to the $\Phi_-$ equation\footnote{This fact was already noticed in \cite{Polchinski:2000uf,Grana:2003ek}.}. Furthermore, as we already mentioned, $\Phi_-$ unambiguously determines the $r^2$ part of the potential.

We therefore conclude that the quadratic piece in the bosonic potential is necessarily of the form:
\begin{equation} \label{Vquad}
V^{\textrm{quad.}}_{\rm D3} =  V_{\mathbf{1}} +
 U_{\mathbf{20}^\prime} \, .
\end{equation}
 We emphasize once more  that the 20 coefficients 
$\mu_{AB}^{\mathbf{20}^\prime}$  in $U_{\mathbf{20}^\prime} $ are added ``by hand" and are fixed only by the boundary conditions. Furthermore, for the $\mathcal{N}=1^\star$ theory ($m_4=0$) we know that this contribution has to be non-zero when the three masses of the chiral multiplets are different. This is obvious from the form of the $\mathcal{N}=1^\star$ bosonic potential in \eqref{eq:V-SUSY}, 
which has terms coming from both the $\mathbf{1}$ (trace) and the $\mathbf{20}^\prime$ representations. Therefore, the solution dual to this theory must contain non-normalizable $L=2$ modes.

We close this section by a short summary: when considering the supergravity dual of the mass-deformed $\mathcal{N}=4$ theory, the backreaction of the fields dual to the fermion masses gives rise to perturbations in the dilaton, metric and 5-form flux proportional to $m_\textrm{f}^2$, but these conspire to yield an overall zero contribution to the traceless part of the quadratic  term of the polarization potential. That term therefore can arise only from the homogeneous  traceless $L=2$ modes that we referred to as $U_{\mathbf{20}^\prime}$. This implies that in order to construct the  supergravity dual of, say, $\mathcal{N}=1^\star$ SYM theory one has to add ``by hand" proper homogeneous $\textbf{20}^\prime$ UV modes in order to ensure that the bosonic masses will match the fermionic ones. 

\section{The trace of the bosonic and fermionic mass matrices}
\label{sec:Future}

From the previous section we can arrive to another crucial observation. From \eqref{Vquad} and the explicit form of the singlet \eqref{pot20} (or \eqref{eq:tilde-V-mm-bar} for a generic mass matrix), we find 
\begin{eqnarray} \label{zerosupertrace}
\text{Tr} [\text{boson\,masses}^2]&=&\text{Tr} [\text{fermion\,masses}^2] \\
\text{Tr} ({\cal M}^2) &=& \text{Tr} (M M^{\dagger})= \text{Tr} (m m^{\dagger}) + 2 \, \hat m_I \bar{\hat m}^I + \tilde m^2  \nonumber \ .
\end{eqnarray}

As explained in the Introduction, this result establishes that only theories where the supertrace of the mass squared is  zero can be described holographically by asymptotically-$AdS$ solutions. The sum of the squares of the boson masses, which is an unprotected operator (also known as {\it the Konishi}) and has been argued to be dual to a stringy mode of dimension $(g_s N)^{1/4}$, can be in fact turned on without turning on stringy corrections, as one could have anticipated from the solutions of \cite{Freedman:1999gp,Pilch:2000ue,Girardello:1999bd}. In the presence of fermion masses, what is dual to a stringy mode is not therefore the sum of the squares of the  boson masses, but rather the mass super-trace (the difference between the sums of the squares of the fermion masses and the boson masses). Theories where this  supertrace is zero can be described without stringy modes, but to describe theories where this supertrace is nonzero, one has to turn on ``stringy'' non-normalizable modes which destroy the $AdS$ asymptotics.

One can also see the relation between this zero-supertrace condition and the existence of an asymptotically-$AdS$ holographic dual from the dual gauge theory. Indeed, in a gauge theory where supersymmetry is broken by adding bosonic masses, there are no quadratic divergences, and the explicit breaking of supersymmetry is called soft. There are other soft supersymmetry-breaking terms that one can add to an ${\cal N}=1$ Lagrangian, such as gaugino masses $\tilde m$, and trilinear bosonic couplings of the form

\begin{equation} \label{LsusyLsoftcubic}
V_{\rm cubic}=   \frac12 c_{IJ}^K \phi^I \phi^J \bar \phi_K + \frac16 a_{IJK} \phi^I \phi^J \phi^K  + h.c.
\end{equation}

Similar to the quadratic terms discussed in the previous section, the bosonic cubic terms can also be read off by considering the action of probe D3 branes. They are proportional to the (3,0) and (2,1) imaginary anti-self-dual piece of the three form flux $T_3$ \cite{Myers:1999ps}, which in turn are determined by the supersymmetry breaking fermionic masses $\hat m_I, \tilde m$ as in (\ref{TandM}) \cite{Grana:2002tu}. One gets\footnote{For exact normalizations see \cite{BBT-antibranes}.}
\begin{equation} \label{ca}
c_{IJ}^K=  \delta^K_{[I} \hat m_{J]} \ , \qquad a_{IJK}=\tilde m \epsilon_{IJK} \ .
\end{equation}

Armed with this knowledge, one can compute the one-loop beta functions for all the coupling constants including the ``non-standard soft supersymmetry breaking" terms $\hat m$ \cite{Jack:1999ud}. If one uses the relation between the soft trilinear terms and the fermion masses (\ref{ca}) we find that all the one loop beta functions except the one for the boson masses vanish exactly \cite{usshort}.  The one-loop beta functions for the boson mass trace vanishes if and only if the  trace of the boson masses is equal to that of the fermions at tree level, which is precisely what happens for the $\mathcal{N}=0^\star$ theories that have an asymptotically-AdS supergravity dual (Eq. \eqref{zerosupertrace}), and also for any gauge theory that has a UV conformal fixed point (such as the ones found on D3 branes at singularities). Since the masses do not run with the scale, this is consistent with the fact that this theory has a UV conformal fixed point. Interestingly enough, the two-loop beta functions \cite{MV,JJ2} also vanish under the same condition \cite{usshort}.

Hence, the field theory computation of the one and two-loop beta functions confirms the results of our holographic analysis: Asymptotically-AdS solutions are dual to theories with UV conformal fixed points, and if one turns on the fermion masses, the sum of the squares of the boson masses is  automatically determined to be equal to the sum of the squares of the fermion masses. Conversely, in perturbative field theory one can turn on arbitrary boson and fermion masses, but for a generic choice of masses the beta-functions will be non-zero and the theory will not have a UV conformal fixed point. These beta-functions only vanish when the sums of the squares of the fermion and boson masses are equal. We can graphically summarize this as two equivalent statements:
\begin{center}
\begin{tabular}{ r c c l }
SUPERGRAVITY: & Asympt-$AdS$ \ $\Leftrightarrow$ \ UV conformal & $\rightarrow$ & $\sum m_\textrm{boson}^2 = \sum m_\textrm{fermion}^2 $ \nonumber \vspace{5pt}\\
FIELD THEORY: & $\sum m_\textrm{boson}^2 \neq \sum m_\textrm{fermion}^2$  & $\rightarrow$ & UV  co$\cancel{\rm{nform}}$al \ $\Leftrightarrow$ Asympt-$\cancel{{AdS}}$ \nonumber
\end{tabular}
\end{center}

\acknowledgments{We would like to thank Johan Bl{\aa}b{\"a}ck, Stefano Massai, Kostas Skenderis, Marika Taylor, David Turton, Nick Warner and Alberto Zaffaroni for insightful discussions. This work was supported in part by the ERC Starting Grants 240210
{\em String-QCD-BH} and 259133  {\em ObservableString}, by the NSF Grant
No.~PHYS-1066293 (via the hospitality of the Aspen Center for Physics)
by the John Templeton Foundation Grant 48222, by a grant from the
Foundational Questions Institute (FQXi) Fund, a donor advised fund of
the Silicon Valley Community Foundation on the basis of proposal
FQXi-RFP3-1321 (this grant was administered by Theiss Research) and by the P2IO LabEx (ANR-10-LABX-0038) in the framework Investissements d'Avenir � (ANR-11-IDEX-0003-01) managed by the French ANR.}

\appendix

\section{'t Hooft symbols}
\label{sec:t-Hooft-matrices}

The explicit form of the 't Hooft matrices $G^A_{ij}$ is

\begin{eqnarray} \label{tHooft}
G^1 = \left(
\begin{array}{cc}
0 & - i \sigma_2 \\
- i \sigma_2 & 0
\end{array}
\right)
\quad
&
\quad
G^2 = \left(
\begin{array}{cc}
0 & - \sigma_0 \\
 \sigma_0 & 0
\end{array}
\right)
\quad
&
\quad
G^3 = \left(
\begin{array}{cc}
i \sigma_2 & 0 \\
 0 & -i \sigma_2
\end{array}
\right)
\nonumber \\
G^4 = \left(
\begin{array}{cc}
0 & - i \sigma_1 \\
i \sigma_1 & 0
\end{array}
\right)
\quad
&
\quad
G^5 = \left(
\begin{array}{cc}
0 & i \sigma_3 \\
 - i \sigma_3 & 0
\end{array}
\right)
\quad
&
\quad
G^6 = \left(
\begin{array}{cc}
\sigma_2 & 0 \\
 0 & \sigma_2
\end{array}
\right)
\end{eqnarray}
Here $\sigma_{1,2,3}$ are the standard Pauli matrices and $\sigma_0$ is the $2\times 2$ unit matrix. These matrices satisfy the following basis independent properties:
\begin{equation}
\label{eq:G-properties-1}
G^A_{ij}  \delta_{A B} {G^B}^{kl} = -2 \left( \delta_i^k \delta_j^l - \delta_j^k \delta_i^l  \right) \ ,
\qquad
\textrm{Tr} \left( {G^A}^\dagger G^B \right) = {G^A}^{ij} {G^B}_{ji}  =  4 \delta_{A B} \, ,
\end{equation}
and
\begin{eqnarray}
\label{eq:G-properties-2}
G_{ik}^A {{G^B}^\dagger}^{kj} + G_{ik}^B {{G^A}^\dagger}^{kj} &=& 2 \delta^{A B} \delta^j_i 
\\ 
i \epsilon_{ABCDEF}{G^A}_{i k_1}  {G^B}^{k_1 k_2} {G^C}_{k_2 k_3}  {G^D}^{k_4 k_5} {G^E}_{k_5 k_6}  {G^F}^{k_6 j}  &=& \delta^j_i \nn .
\end{eqnarray}

\bibliographystyle{utphys}
\bibliography{Draft}

\end{document}